\documentclass[aps,prb,twocolumn,groupedaddress,showpacs,floatfix]{revtex4}
\usepackage{graphicx}
\usepackage{epsfig}
\usepackage{amsmath}
\usepackage{amssymb}

\begin{document}


\title{Charge-ice dynamics in the negative thermal expansion material Cd(CN)$_2$}

\author{Vanessa E. Fairbank}
\affiliation{Department of Chemistry, University of Oxford, Inorganic Chemistry Laboratory, South Parks Road, Oxford OX1 3QR, U.K.}

\author{Amber L. Thompson}
\affiliation{Department of Chemistry, University of Oxford, Inorganic Chemistry Laboratory, South Parks Road, Oxford OX1 3QR, U.K.}

\author{Richard I. Cooper}
\affiliation{Department of Chemistry, University of Oxford, Inorganic Chemistry Laboratory, South Parks Road, Oxford OX1 3QR, U.K.}

\author{Andrew L. Goodwin}
\email[]{andrew.goodwin@chem.ox.ac.uk}
\affiliation{Department of Chemistry, University of Oxford, Inorganic Chemistry Laboratory, South Parks Road, Oxford OX1 3QR, U.K.}

\date{\today}
\begin{abstract}
We use variable-temperature (150--300\,K) single-crystal X-ray diffraction to re-examine the interplay between structure and dynamics in the ambient phase of the isotropic negative thermal expansion (NTE) material Cd(CN)$_2$. We find strong experimental evidence for the existence of low-energy vibrational modes that involve off-centering of Cd$^{2+}$ ions. These modes have the effect of increasing network packing density---suggesting a mechanism for NTE that is different to the generally-accepted picture of correlated Cd(C/N)$_4$ rotation modes. Strong local correlations in the displacement directions of neighbouring cadmium centres are evident in the existence of highly-structured diffuse scattering in the experimental X-ray diffraction patterns. Monte Carlo simulations suggest these patterns might be interpreted in terms of a basic set of `ice-rules' that establish a mapping between the dynamics of Cd(CN)$_2$ and proton ordering in cubic ice VII.
\end{abstract}


\maketitle

\begin{figure}
\includegraphics{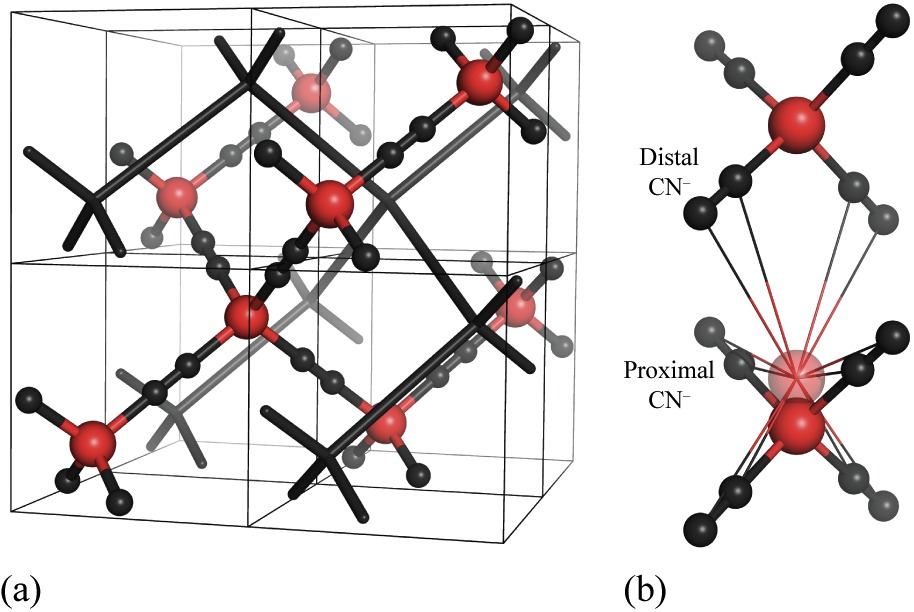}
\caption{\label{fig1} (a) The $Pn\bar3m$ anticuprite structure of Cd(CN)$_2$: Cd centres are shown as large red spheres, and cyanide ions as small black spheres. The cyanide orientations are disordered; consequently no distinction is made between C and N atoms. Like cuprite itself, two topologically independent networks interpenetrate. Here one is shown in ball-and-stick representation and the other in stick representation. (b) Cadmium off-centering in Cd(CN)$_2$. Displacement of the Cd$^{2+}$ ions parallel to one of the $\langle100\rangle$ crystal axes gives rise to a pseudo-octahedral coordination environment.}
\end{figure}

Of the few materials known to exhibit the rare property of isotropic negative thermal expansion (NTE; volume contraction on heating), cubic Cd(CN)$_2$ shows the strongest such effect.\cite{Goodwin_2005,Barrera_2005,Phillips_2008} Its volume coefficient of thermal expansion $\alpha={\rm d}\ln V/{\rm d}T=-61.2(12)$\,MK$^{-1}$ is substantially more negative than the thermal expansion coefficients of typical engineering materials are positive ($\alpha\simeq30$\,MK$^{-1}$).\cite{Okaji_2000} The origin of NTE in this material has never been identified conclusively, despite the related material Zn(CN)$_2$ (which also shows NTE, but at a reduced level) having been the focus of a significant number of experimental and computational studies.\cite{Goodwin_2005,Chapman_2005,Ravindran_2007,Zwanziger_2007,Poswal_2009,Mittal_2011} Both compounds share the anticuprite structure in which tetrahedral Zn or Cd centres are connected via linear M--C--N--M linkages [Fig.~\ref{fig1}(a)].\cite{Shugam_1945,Iwamoto_1997} Density functional theory (DFT) calculations suggest that the phonon relations are similar in the two materials, with the more pronounced NTE effect of Cd(CN)$_2$ a result of an enhanced pressure sensitivity of the low-energy mode frequencies.\cite{Zwanziger_2007} The atomic displacements associated with these crucial low-frequency modes are generally assumed to involve concerted rotations of M(C/N)$_4$ coordination tetrahedra, which result in transverse displacements of the C and N atoms away from the M$\ldots$M axes.\cite{Goodwin_2005,Chapman_2005,Goodwin_2006,Zwanziger_2007,Poswal_2009,Mittal_2011} Thermal population of these modes would give rise to larger transverse displacements---hence reducing both the M$\ldots$M separation and the unit cell dimensions.

Yet there are reasons to suspect that the dynamical behaviour in Cd(CN)$_2$ and Zn(CN)$_2$ might differ more strongly than is presently appreciated. It is known, for example, that Cd(CN)$_2$ exhibits a structural phase transition at 150\,K that is not observed in Zn(CN)$_2$ at temperatures down to 20\,K (albeit that the precise nature of this transition has not yet been reported).\cite{Goodwin_2005} This means that the $Pn\bar3m$ anticuprite structure used in DFT calculations\cite{Zwanziger_2007} does not actually reflect the thermodynamic ground state at 0\,K. And whereas tetrahedral geometry is the most common coordination motif for Zn$^{2+}$ ions in cyanide complexes,\cite{Hoskins_1990,Hoskins_1995} the larger Cd$^{2+}$ cation ($r\simeq0.95$\,\AA\ for Cd$^{2+}$ \emph{vs} 0.60\,\AA\ for Zn$^{2+}$)\cite{Shannon_1976} nearly always adopts an octahedral coordination environment in otherwise equivalent compounds.\cite{Sharpe_1976,Hoskins_1994,Iwamoto_1997} This difference in geometry preference is relevant because it has the potential to influence the relative importance of geometry-preserving `rigid-unit modes' (RUMs)---precisely the type of phonon mode usually implicated in NTE for these materials.\cite{Goodwin_2005,Goodwin_2006,Zwanziger_2007}

In this study, we use variable-temperature (150--300\,K) single-crystal X-ray diffraction to re-examine the interplay between structure and dynamics in the ambient-temperature phase of Cd(CN)$_2$. We find strong experimental evidence for low-energy vibrational modes that involve off-centering of Cd$^{2+}$ ions. These modes have the effect of increasing network packing density---so providing an alternative mechanism for NTE. Strong local correlations in the displacement directions of neighbouring cadmium centres are evident in the existence of highly-structured diffuse scattering in the experimental X-ray diffraction patterns. With the aid of Monte Carlo simulations we are able to interpret these patterns in terms of a basic set of `ice-rules' that establish a mapping between the dynamics of Cd(CN)$_2$ and proton ordering in cubic ice VII.

Our study makes use of a single crystal of Cd(CN)$_2$ that was prepared by slow evaporation of a concentrated aqueous solution (see SI for details). X-ray diffraction patterns were collected using an Oxford Diffraction (Agilent) SuperNovae diffractometer covering a full sphere of reciprocal space, with the same experimental procedure repeated at 30\,K intervals between 150 and 300\,K; refinements were carried out with CRYSTALS\cite{Betteridge_2003} (see SI for further experimental details). While highly-structured diffuse scattering was clearly evident in each of these six data sets, the observed Bragg reflection conditions were consistent with the $Pn\bar3m$ structure reported in previous studies.\cite{Goodwin_2005} On cooling below 150\,K the bulk of this diffuse scattering intensity transferred into a large number of additional superlattice Bragg reflections (although some diffuse scattering remained). The positions of these additional reflections suggested an initial doubling, then quadrupling of the unit cell in one or more directions. The complexity of this diffraction pattern and the likelihood of complex twinning associated with the phase transition means that we have restricted the scope of this particular study to the higher-symmetry ambient phase.

\begin{figure}
\includegraphics{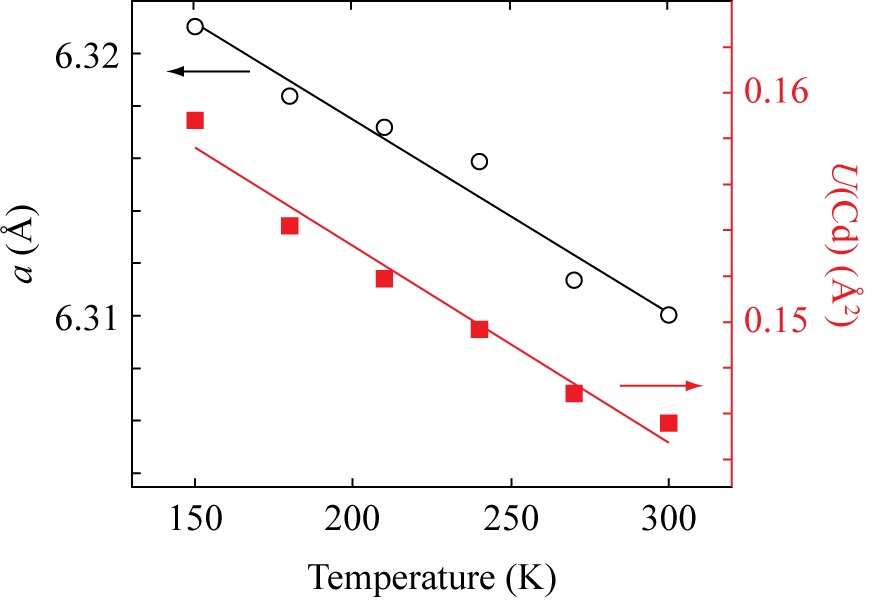}
\caption{\label{fig2} Temperature-dependent structural evolution in Cd(CN)$_2$ as determined using single-crystal X-ray diffraction. With increasing temperature, both the cubic lattice parameter $a$ (open black circles) and the Cd mean-squared displacement $U$(Cd) (filled red squares) decrease. The corresponding linear fits to data are included as guides-to-the-eye; estimated standard errors are smaller than the data points.}
\end{figure}

The cubic unit cell parameters extracted from our diffraction data reflect the previously-reported NTE effect [Fig.~\ref{fig2}].\cite{Goodwin_2005} There are relatively few structural parameters to be refined crystallographically; a summary of the various values we obtain across the entire temperature series is given in Table S1. Here we focus on one crucial parameter: namely, the isotropic displacement parameter of the Cd atom, $U$(Cd). This parameter describes the mean-squared displacement of the Cd atom away from its crystallographic site, and is well constrained by X-ray diffraction intensities by virtue of the large $Z$ contrast between Cd and C/N atoms.\cite{notecdsite} Counterintuitively, we find that its magnitude actually decreases with increasing temperature [Fig.~\ref{fig2}]. Working within the high-temperature approximation, we have
\begin{equation}
U({\textrm{Cd}})=\langle u^2\rangle=\frac{k_{\textrm B}T}{m\omega_{\textrm{eff}}^2},
\end{equation}
where $m$ is the mass of Cd and $\hbar\omega_{\textrm{eff}}$ an effective (average) phonon energy. Consequently, a negative value of ${\textrm d}U/{\textrm d}T$ implies a large and positive value of ${\rm d}\omega_{\textrm{eff}}/{\rm d}T$---\emph{i.e.}, the existence of strongly anharmonic low-frequency phonon modes, with the corresponding mode eigenvectors involving Cd displacements. Such behaviour is commonplace in NTE materials, with the corresponding atomic displacement patterns generally interpreted as the microscopic driving force for NTE itself .\cite{Ernst_1998,Ramirez_1998,Cao_2002,Drymiotis_2004,Chapman_2006} So the fact that Cd displacements have a strong and anomalous temperature dependence in Cd(CN)$_2$ suggests that these displacements are crucial to the NTE mechanism. We note that the rotational RUMs discussed in the introductory paragraph, and cited elsewhere as a possible driving force for NTE in this material,\cite{Goodwin_2006,Zwanziger_2007} do not give rise to Cd displacements.

\begin{figure}
\includegraphics{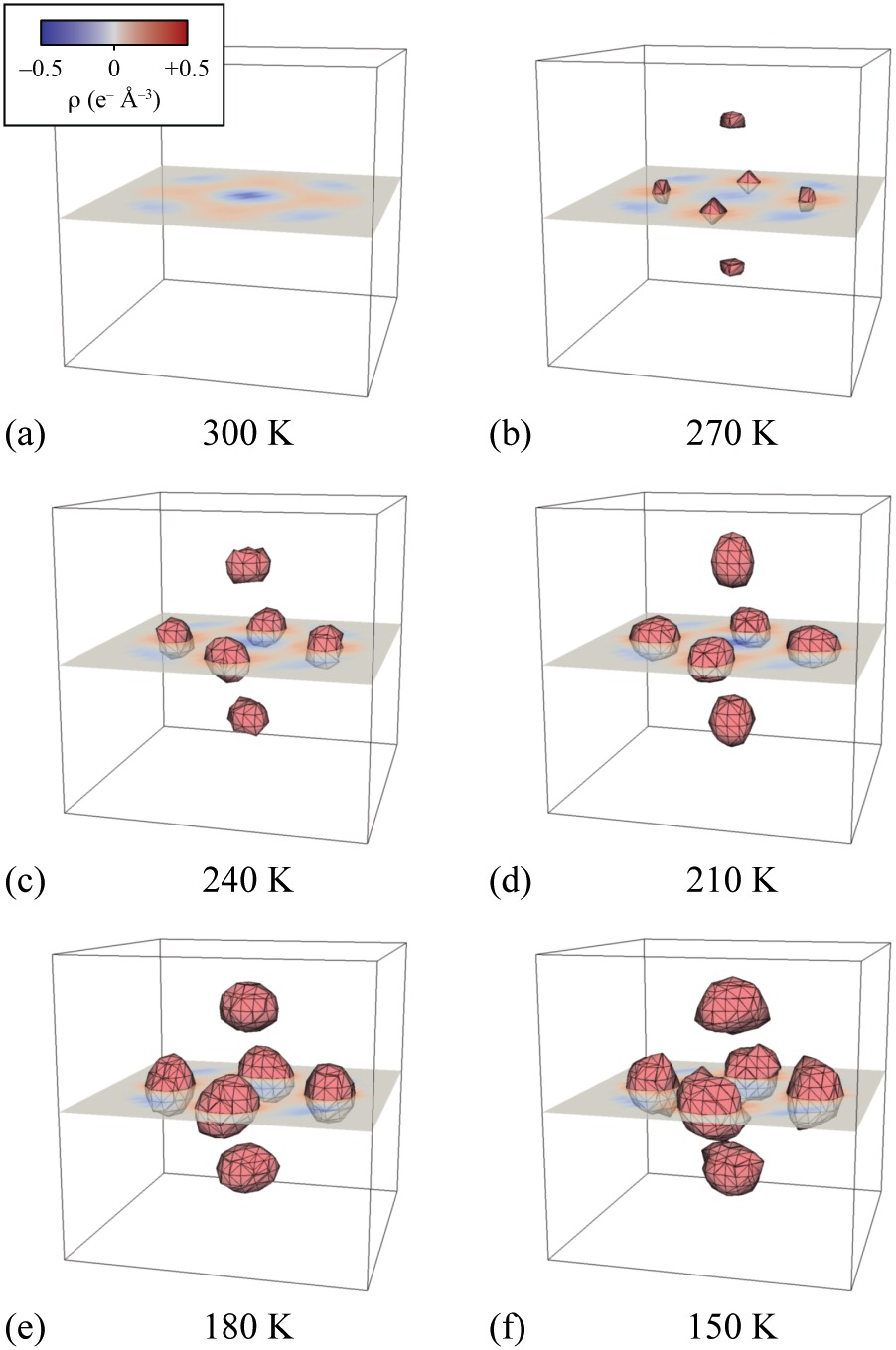}
\caption{\label{fig3}Thermal evolution of residual scattering density around the Cd site in the difference Fourier map for Cd(CN)$_2$. Each cube corresponds to an octant of the unit cell $0\leq(x,y,z)\leq\frac{1}{2}$, with the $(\frac{1}{4},\frac{1}{4},\frac{1}{4})$ Cd position at its centre and the $\langle100\rangle$ crystal axes parallel to the cube edges. Isosurfaces are drawn at a level of $\rho=+0.2$\,e$^-$\AA$^{-3}$.}
\end{figure}

In order to investigate more fully the form of these displacement modes, we calculated difference Fourier maps for the Cd site for each of the single crystal X-ray diffraction data sets at hand; the corresponding functions are illustrated in Fig.~\ref{fig3}. At 300\,K the modulations in the difference function are small, indicating that there are no especially important regions of the unit cell where electron density has neither been sufficiently accounted for nor overaccounted. For data sets collected at successively lower temperatures, a set of `lobes' of unaccounted-for electron density emerges, with these lobes positioned at approximately 1\,\AA\ away from the Cd site in each of the six symmetry-equivalent $\langle100\rangle$ directions. The picture that emerges resembles closely the behaviour observed in archetypal dynamic order--disorder transitions, such as the displacement of Pb atoms along $\langle100\rangle$ axes in cubic PbTiO$_3$.\cite{Nelmes_1990}

Here in Cd(CN)$_2$, displacements along the $\langle100\rangle$ directions have an intriguing effect on the coordination geometry of the Cd$^{2+}$ ion. The displacement is towards an edge of the Cd(C/N)$_4$ coordination tetrahedron, bringing the Cd centre closer to two of its four cyanide ligands (we refer to these as `proximal' CN$^-$ ions), and further away from two others [Fig.~\ref{fig1}(b)]. In addition, a new association is now made with a further two cyanide ions belonging to the adjacent Cd(CN)$_2$ network (we refer to these as `distal' CN$^-$ ions). Consequently, the Cd off-centering we observe reflects a drive towards increased Cd coordination number: \emph{i.e.}\ from four to six. At the same time, the cyanide ions undergo an effective increase in coordination number from two towards three as they are now associated with an additional Cd$^{2+}$ in a sideways geometry [Fig.~\ref{fig1}(b)]. The strong X-ray scattering contrast between Cd and C/N atoms means that we are not especially sensitive to C and N displacements; however we note that there is clear evidence for transverse displacements of the distal cyanide ions towards the approaching Cd$^{2+}$ centre in the difference Fourier maps (see SI), suggesting possible anharmonic coupling of Cd translations along $\langle100\rangle$ directions with transverse CN displacements.

Because thermal excitation of these modes increases the packing density (the unperturbed $Pn\bar3m$ geometry being maximally extended), each mode will be associated with a negative  Gr{\"u}neisen parameter $\gamma={\rm d}\ln\omega/{\rm d}\ln V<0$---hence a driving force for NTE.\cite{Barrera_2005} Indeed a similar displacement pattern is observed in reverse as a mechanism for framework \emph{expansion} in rutile-TiO$_2$ under \emph{negative pressure}: the (6,3)-net expands at $-10$\,GPa to give a lower-density distorted tetrahedral TiO$_2$ net with the anticuprite topology.\cite{Liu_2009} Distortive modes have also been suggested as a possible partial mechanism for NTE in the related framework NMe$_4$[CuZn(CN)$_4$], although in that system the displacements involved are much smaller ($<0.05$\,\AA).\cite{Phillips_2010}

\begin{figure}
\includegraphics{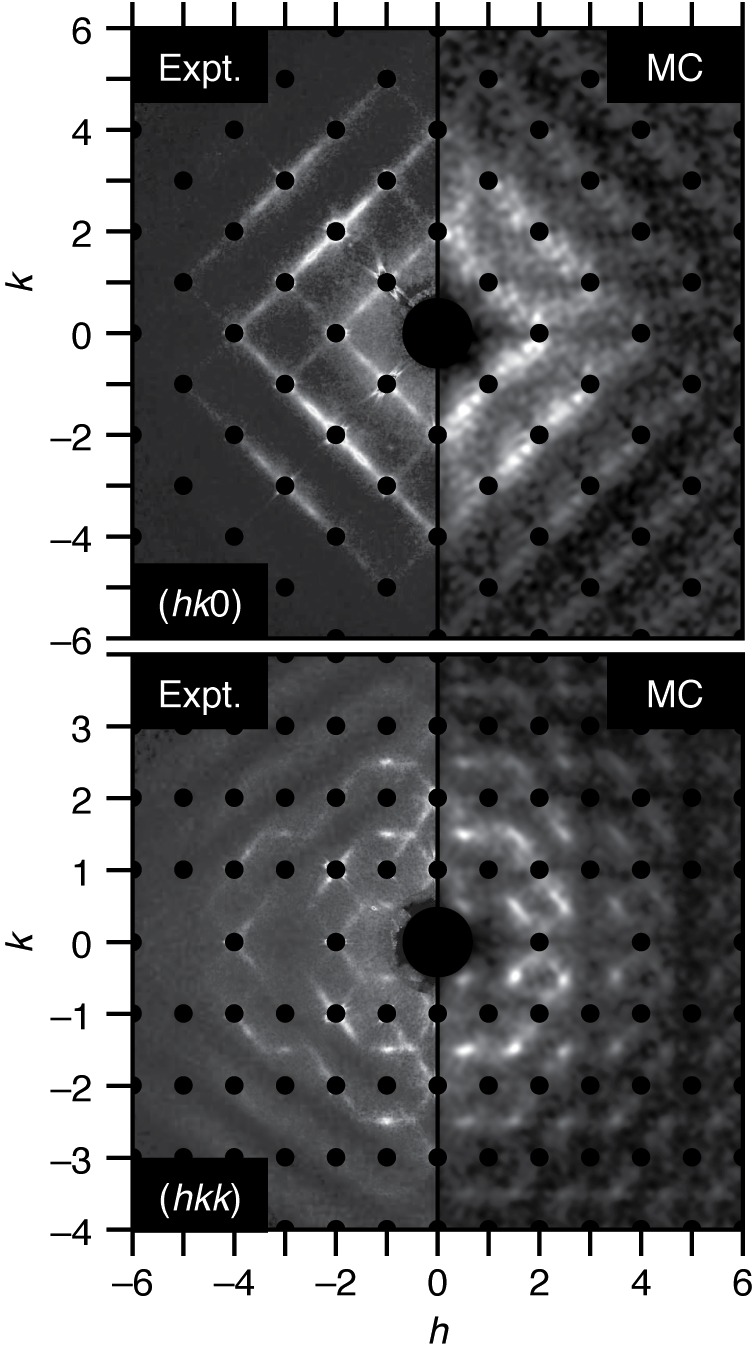}
\caption{\label{fig4}(Left) Experimental X-ray diffuse scattering patterns collected at 270\,K for the $(hk0)$ and $(hkk)$ cuts of reciprocal space and (right) the corresponding predicted diffuse scattering from the Monte Carlo model as described in the text.}
\end{figure}

A displacement of \emph{ca} 1\,\AA\ is sufficiently large that one might anticipate strong correlation in displacement directions for neighbouring Cd atoms. Indeed, the existence of such local correlations is supported experimentally by the observation of highly structured X-ray diffuse scattering [Fig.~\ref{fig4} and S1--S3]. The intensity of this diffuse scattering is temperature-dependent (see SI), indicating that its origin is dynamic rather than either static or compositional.\cite{Withers_2005} The scattering is also transverse polarised in that the intensity of the observed diffuse streaking is strongest along directions of reciprocal space perpendicular to the streaking itself; this is a second indication that its origin is displacive.\cite{Withers_2005,Goodwin_2007} In general terms the scattering takes the form of diffuse planes, running across the $\langle hkl\rangle^\ast\pm\epsilon\langle h+l,\bar h+l,\bar{2l}\rangle^\ast$ regions of reciprocal space. It also obeys the extinction conditions $h+k+l=2n$, allowing us to infer that the atoms responsible are separated by $\frac{1}{2}\langle 111\rangle$ vectors in real-space:\cite{Withers_2005} \emph{i.e.}, the Cd$\ldots$Cd separation. For completeness, we note that while there are likely to be correlations between cyanide ion displacements, the large contrast in Cd \emph{vs} C/N scattering by X-rays will mean that diffuse scattering due to Cd displacements alone dominates the experimental data.

\begin{figure}
\includegraphics{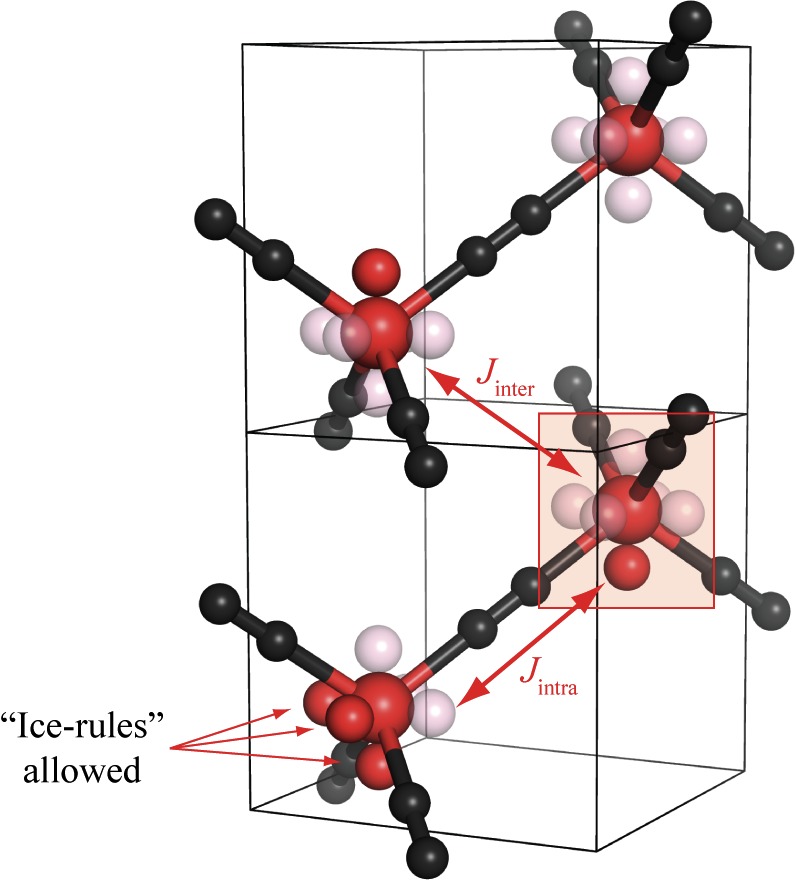}
\caption{\label{fig5}Correlated Cd$^{2+}$ displacements in Cd(CN)$_2$. A requirement that each CN$^-$ be proximal to a single Cd$^{2+}$ centre implies ice-rules-like correlations in Cd$^{2+}$ displacements. Here, a displacement of one reference Cd$^{2+}$ centre (highlighted) in a downwards direction reduces the number of possible displacement directions from six to three for the Cd$^{2+}$ centre to which it is immediately connected within its covalent network. Of these three ice-rules-allowed directions, one is parallel to the original displacement vector (ferroelectric displacements), and two are perpendicular. Within MC simulations, the coupling parameter $J_{\textrm{intra}}$ can be used to favour either type of correlated displacement. Similarly, \emph{inter}-network correlations can be controlled through the coupling parameter $J_{\textrm{inter}}$. The calculated diffuse scattering patterns shown in Fig.~\ref{fig4} were generated using a pair of $J_{\textrm{inter}},J_{\textrm{intra}}$ values that favour ferroelectric displacements within the same network, but antiferroelectric displacements between networks.}
\end{figure}

So what is the simplest possible ordering model for Cd displacements that could explain the general form of this diffuse scattering? Our starting point in answering this question was the single assumption that the close approach between a given Cd$^{2+}$ centre and its proximal cyanides [Fig.~\ref{fig1}(a)] means that any cyanide ion is likely to be proximal only to a single Cd$^{2+}$ centre. In this way, the corresponding Cd--CN bond might relax to accommodate the initially-decreased Cd$\ldots$C/N separation. It transpires that there exists a one-to-one mapping between this rule and the problem of proton ordering in cubic ice: just as each O atom in the latter is associated more closely with two of its four H neighbours, so too is each Cd centre closer to two of its four cyanide neighbours in Cd(CN)$_2$. And, just as each proton in cubic ice is covalently bonded to a single O atom, each cyanide ion in Cd(CN)$_2$ is proximal only to a single Cd$^{2+}$ centre. Consequently, if this rule were to hold then for any given pair of connected Cd atoms their displacements must be `ice-rules' allowed [Fig.~\ref{fig5}].

Making use of a simple Monte Carlo (MC) routine, we generated Cd(CN)$_2$ atomistic configurations with Cd centres displaced according to this rule and subsequently calculated the corresponding single-crystal diffuse scattering patterns with the approach described in Refs.~\onlinecite{Butler_1992,Goodwin_2007} (see SI for details). We find that---on its own---this ice-rules-only algorithm predicts modulated diffuse scattering that is not yet convincingly close to the observed patterns [Fig.\ S5]. Consequently, our next step was to increase the complexity of the MC algorithm by considering explicitly the correlations between the directions of nearest-neighbour Cd displacements. There are two types of Cd$\ldots$Cd pairs separated by $\frac{1}{2}\langle111\rangle$ vectors: `intraframework' pairs connected via cyanide ions, and `interframework' pairs which are not directly connected at all. By calculating diffuse scattering patterns for a range of different MC coupling constants $J_{\rm{inter}},J_{\rm{intra}}$ we found the strongest resemblance to the experimental data for a model where intra-framework pairs displace in a like sense, and inter-framework pairs displace in opposite senses. Two representative diffuse scattering patterns are shown in Fig.~\ref{fig4}, and full details of the various models are given as SI.

The dominant Cd displacement mode is consequently an antiferroelectric (zone boundary) mode, corresponding to a local symmetry lowering from the space group $Pn\bar3m$ to $I4_1/amd$: the displacements within either one of the two interpenetrated networks tend to occur in similar directions, but the direction is reversed from one network to the other. The distorted $I4_1/amd$ ($\sqrt2\times\sqrt2\times2$) cell may well provide a useful starting point for structure solution of the low-temperature phase. Intriguingly, the same space-group relation is found for the ice VII/VIII transition.\cite{Kuhs_1984} While the role of static disorder in cubic ice VII remains a controversial and difficult problem,\cite{Nelmes_1998,Nelmes_1993} it is pertinent to note that the phase is thought to exhibit isotropic NTE for some regions of its stability field.\cite{Frank_2004} As for ice itself, one anticipates that Cd(CN)$_2$ must exhibit a large entropic signature associated with Cd displacement ordering at the 150\,K phase transition. We have not attempted to measure the temperature dependence of the specific heat as part of this study, noting that the added complication of CN orientational disorder may render specific heat anomalies difficult to interpret.\cite{Nishikiori_1990} 

Returning to the displacement pattern shown in Fig.~\ref{fig1}(b), we note that the antiferroelectric modes we have identified will mean that pairs of unconnected Cd$^{2+}$ centres displace in opposite senses, giving rise to localised polar dimers (effectively a Cd$_2$ quadrupole). That is, the proximal cyanides of one Cd$^{2+}$ ion become the distal cyanides of the corresponding inter-framework Cd$^{2+}$ ion and \emph{vice versa}. Very similar behaviour has been observed in frustrated niobate pyrochlores such as Y$_2$NbTiO$_7$ and Y$_2$Nb$_2$O$_7$, where correlated Nb off-centering is thought to be driven by the propensity of Nb$^{4+}$ to form bonded metal--metal dimers.\cite{McQueen_2008,Blaha_2004} Whereas the Nb--Nb contacts are fixed and hence the systems represent a static `charge ice', in cubic Cd(CN)$_2$ the charge-ice displacements are clearly dynamic.

In summary, we have shown that Cd off-centering is a dominant dynamical process in the NTE material Cd(CN)$_2$. Displacements of the Cd$^{2+}$ ions away from their high-symmetry (crystallographic) site have the effect of increasing their coordination number and also result in a contraction of the cubic unit cell. We find that experimental X-ray diffraction patterns for Cd(CN)$_2$ contain highly-structured diffuse scattering. This diffuse scattering, which has a dynamical origin, can be interpreted in terms of correlated displacements of Cd$^{2+}$ centres according to a set of local antiferroelectric `ice-rules'. The existence of charge-ice dynamics in Cd(CN)$_2$ lends further weight to the emerging correspondence between geometric frustration and NTE:\cite{Cao_2002,Hancock_2004,Ramirez_2000} the same configurational degeneracy responsible for entropic anomalies in cubic ice is also implicated in driving NTE in Cd(CN)$_2$.

\section*{Acknowledgements}

The authors gratefully acknowledge valuable discussions with J.~A.~M.~Paddison, C.~J.~Kepert, D.~A.~Keen and M.~G.~Tucker. We are grateful for financial support from the E.P.S.R.C. (Grant No. EP/G004528/2) and the E.R.C. (Grant No. 279705).


\begin{references}
\bibitem{notecdsite}
The site symmetry at the Cd position (namely, $\bar43m$) is such that the atomic displacement tensor $U_{ij}$ is intrinsically isotropic.
\end{references}
\end{document}